\documentclass[fleqn,10pt]{wlscirep}
\usepackage[utf8]{inputenc}
\usepackage[T1]{fontenc}
\usepackage{pdfpages}
\usepackage{lineno}
\usepackage{tabulary}
\usepackage{scalerel}
\usepackage{tikz}
\usetikzlibrary{svg.path,calc}

\definecolor{orcidlogocol}{HTML}{A6CE39}
\tikzset{
   orcidlogo/.pic={
    \fill[orcidlogocol] svg{M256,128c0,70.7-57.3,128-128,128C57.3,256,0,198.7,0,128C0,57.3,57.3,0,128,0C198.7,0,256,57.3,256,128z};
    \fill[white] svg{M86.3,186.2H70.9V79.1h15.4v48.4V186.2z}
                 svg{M108.9,79.1h41.6c39.6,0,57,28.3,57,53.6c0,27.5-21.5,53.6-56.8,53.6h-41.8V79.1z M124.3,172.4h24.5c34.9,0,42.9-26.5,42.9-39.7c0-21.5-13.7-39.7-43.7-39.7h-23.7V172.4z}
                 svg{M88.7,56.8c0,5.5-4.5,10.1-10.1,10.1c-5.6,0-10.1-4.6-10.1-10.1c0-5.6,4.5-10.1,10.1-10.1C84.2,46.7,88.7,51.3,88.7,56.8z};
  }
}

\newcommand\orcidicon[1]{\href{https://orcid.org/#1}{\mbox{
\begin{tikzpicture}[overlay,remember picture]
\coordinate (A);
\coordinate(B) at ($(A)-(2pt,-9pt)$);
\end{tikzpicture}
\begin{tikzpicture}[overlay,remember picture,yscale=-0.0381,xscale=0.0381,transform shape]
\pic at (B) {orcidlogo};
\end{tikzpicture}
}{}}}

\usepackage{hyperref}
 
\title{AMC 12 atomic mass compilation data extrapolated for atomic masses of nuclei far from the valley of stability
}

\author[1,2]{K. Venkataramaniah}
\author[2,*]{Shreesha Rao D. S.\protect\orcidicon{0000-0003-2007-0930}\,\,}
\author[1]{C. Scheidenberger}

\affil[1]{GSI Helmholtzzentrum f\"{u}r Schwerionenforschung GmbH, 64291, Darmstadt, Germany}
\affil[2]{Department of Physics, Sri Sathya Sai Institute of Higher Learning, Prasanthinilayam, Anantapur, 515134, India.}

\affil[*]{\small corresponding author: Shreesha Rao D. S. (shreesharaods@gmail.com)}

\begin{abstract}
The experimental mass data from the Atomic Mass Compilation - 2012 (AMC12) has been analyzed for two-neutron separation energies (S$_{2n}$), two-proton separation energies (S$_{2p}$), double-beta decay energies (Q$_{2\beta^-}$), and four-beta decay energies (Q$_{4\beta^-}$) and plotted against neutron number and mass number, respectively. A new weighted slope method of extrapolation, tested for known and new mass measurements, has been used to obtain the extrapolated mass values with better precision for more than 1100 nuclei far from the valley of stability, out of which more than 100 are being reported for the first time. A comparison has been made with five of the popular mass models with reference to experimental extrapolated masses from the present work and the Atomic Mass Evaluation 2016 (AME16). The extrapolated experimental atomic mass data will be very useful for both experimentalists and mass-model theoreticians, as well as in simulations of astrophysical r-processes. 
\end{abstract}

\begin{document}
\flushbottom
\maketitle

\section*{Background \& Summary}
The fundamental characteristics of an atomic nucleus can be understood by knowing its mass. The atomic mass value determines the binding energy of the atom and allows us to establish the decay rates, reactions, and stability of the atom. The nucleosynthesis in stars has been studied in order to better understand the formation of elements as well as the abundance of some of the elements in the universe. The synthesis of a nucleus is influenced by environmental factors like temperature and neutron density as well as by the characteristics of the nuclides themselves. Understanding and modelling these astrophysical processes requires a precise knowledge of the constituent atomic masses. These astrophysical processes, however, take place far from the valley of stability, where experimental masses are not known, and no theoretical model exists that can precisely predict these atomic masses. Therefore, in the absence of any experimental atomic mass measurements in this region and reliable model-based extrapolations, precise local extrapolations of the available experimental atomic mass data would be of great help in arriving at the atomic masses of nuclei far from the valley of stability. Furthermore, atomic mass data from nuclei far from the valley of stability would shed light on the evolution of shell closures, nucleon-nucleon correlations, fundamental symmetries, and nuclear existence limits. They will also aid in the improvement of nuclear models. 

Precise knowledge of nuclear masses in regions away from measured values is an important topic in nuclear physics~\cite{RevMod03Lun}, but many of these masses remain unknown. The values of the unknown nuclear masses are determined by theoretical calculations. However, there is a lack of agreement, and the majority of estimates disagree greatly with one another, particularly in the region of large neutron excess~\cite{RevMod03Lun,PhyRep06Bla}. The models and techniques to predict nuclear masses~\cite{RevMod03Lun} can essentially be classified into global and local approximations.

Global models make use of all experimentally measured nuclear masses. Examples of these kinds of models are the liquid drop model~\cite{NuPy96MyNuP}, the finite-range droplet model (FRDM 12), where the calculations are based on the finite-range droplet macroscopic, and the folded-Yukawa single-particle microscopic nuclear-structure model~\cite{NuG16MolAdndt,Nug95MolAdndt}. The Duflo-Zuker model~\cite{Mic95PhRDuf} assumes the existence of a pseudo potential smooth enough to do Hartree-Fock variations and good enough to describe nuclear structure and construct mass formulas. The Hartree-Fock-Bogoliubov (HFB) mass models~\cite{FurExp10PhRGor}, labeled as HFB-19, HFB-20, and HFB-21, use unconventional Skyrme forces containing density-dependent generalizations. In the infinite nuclear matter (INM) model~\cite{MaP12NayAdndt}, the classical liquid drop is replaced by an INM sphere characterizing the interacting many-fermionic liquid and the model~\cite{NuG16MolAdndt,FurExp10PhRGor,Gh83PrlSat} of many-body theory, which therefore takes into account the shell and the liquid-drop features non-perturbatively. KUTY~\cite{Nuc05PtpKou} is a nuclidic mass formula. This is a revised version of the mass formula constructed by Koura et al. and published in 2000~\cite{Nuc00NpaHir}. The WS4~\cite{Sur14PlbWan} is the most recent and improved version of the semi-empirical nuclear mass formula based on the macroscopic method together with the Skyrme energy-density functional, the surface diffuseness correction for unstable nuclei, and the radial basis function (RBF) corrections. These global mass models can be regarded as one of the methods for determining the masses of experimentally not (yet) accessible nuclides.

Local interpolations or extrapolations of the experimental mass values have been tried as an alternative and more plausible way to reach the nuclides far from the valley of stability. The Isobaric Multiplet Mass Equation (IMME) is a powerful interpolation/extrapolation method for lighter nuclei. Mass values derived from IMME had been included in older mass evaluations and the recent high-precision mass measurements have sparked renewed interest in this technique~\cite{Iso13AdndtHua,Eva14NpaMac}.

On the basis of the semi-empirical Bethe-Weizsacker~\cite{BDe54PhRKat,Zur35WeiZfp} mass formula, Way and Wood~\cite{NpA36RmpBet} derived graphical presentations of nuclear decay data. Under the assumption that the atomic masses are smoothly varying functions of neutron number (N) and atomic number (Z), as in the semi-empirical mass formula, the lines were extrapolated to unmeasured nuclei. In 1966, Garvey and Kelson presented a relationship based on an independent particle model of the nucleus~\cite{NeN66PrlGar}:
\begin{equation} \label{Eq:Eq1}
M (N + 2, Z - 2) - M (N, Z) + M (N, Z - 1) - M (N + 1, Z - 2) + M (N + 1, Z)- M (N + 2,Z - 1) = 0
\end{equation}
This equation relates to six nuclides. With five known masses, a sixth unknown mass can be predicted~\cite{Mag18PrlMic}. These global mass formulas, however, will in general be too approximate to make predictions to a sufficient degree of accuracy.

Audi-Wapstra successfully employed a local extrapolation method~\cite{Amt71AdndtWap} based on the systematics and smoothness of the mass surface and its derivatives. These mass measurements and extrapolations were provided as Atomic Mass Evaluations (AME) and have been regularly updated. An interactive graphical program was used to make estimates for unknown masses from the trends in the mass surfaces. However, this procedure encompasses a ``subjective'' component in the form of individual judgments.

In an attempt to provide extrapolated masses complementary to the extrapolations in AME, we have used an independent data set of mass values, namely the Atomic Mass Compilation 2012 (AMC12)~\cite{Amc14AnndtPfe}. There are numerous potential applications for the AMC12. Predicting accurate estimations of unknown, imperfectly known, or ambiguous masses; extrapolation toward the drip lines; and testing theoretical models are some of these key applications. Such an attempt has been made in this work. One of our aims in using an independent data set rather than AME mass compilations and a new extrapolation method is that the presence of two complementary, but independent sets of data would boost the confidence in using these extrapolated mass estimates. In addition to this, following the tradition of providing extrapolated masses of unknown masses for nuclides far from stability by the Atomic Mass Evaluations (AME), having published the Atomic Mass Compilation (AMC12), we wanted to provide extrapolated masses based on the AMC12 mass data as an extension of AMC12 work using a different extrapolation method.

We have carried out a point-to-point local extrapolation to systematically take into account the local variation in the trends that are observed in the derivative sheets. The local trend was quantified in terms of the value of the slopes connecting one isotope to the next. The extrapolated slope for the prediction was obtained by averaging the slopes of three of the immediate neighbors. A weighted average of the three slopes was used, with higher weightage placed on the slope closest to the one to be found. The extrapolated slope value was used to find the unknown value of the derivative. This extrapolation method was independently carried out on four derivative sheets. The error in the extrapolated derivative was calculated by considering the error in the value of the derivatives used for the individual extrapolations. Since our extrapolation method is based on systematic local extrapolation, the method is very effective in following local variations in trends present in the derivative sheets. The limitation of the method would be the one where there are conflicting slopes while averaging the three slopes (i.e., absence of trend). As can be observed from the derivative sheets (shown below), consistent trends are observed while plotting individual values. Hence, our method succeeds in accurately predicting the unknown mass values.
\section*{Methods}
Plotting nuclear masses as a function of N and Z produces a surface in three dimensions. Due to pairing energy, this surface is separated into four sheets that run parallel in all directions~\cite{Amt71AdndtWap}. This regularity can also be observed in the derivative sheets. A derivative is any specified difference in the masses of two adjacent nuclei. Remarkably smooth trends can be observed in the derivate sheets and this regularity can be used for obtaining unknown masses of nuclides by extrapolation from the well-known mass values on the same derivative sheet. The derivative sheets have the advantage of displaying much smaller variations.
\subsection*{Analysis of experimental Atomic Mass data from AMC12.} 
The simple trends in the mass surfaces have been used to obtain unknown masses by observing the regularities in derivatives mentioned below. We consider four different mass-sheets: S$_{2n}$ as a function of neutron number, S$_{2p}$ as a function of proton number, Q$_{2\beta^-}$ as a function of mass number and, Q$_{4\beta^-}$ as a function of mass number.

\begin{figure}[htbp!]
\centering
\includegraphics[width=0.9\linewidth]{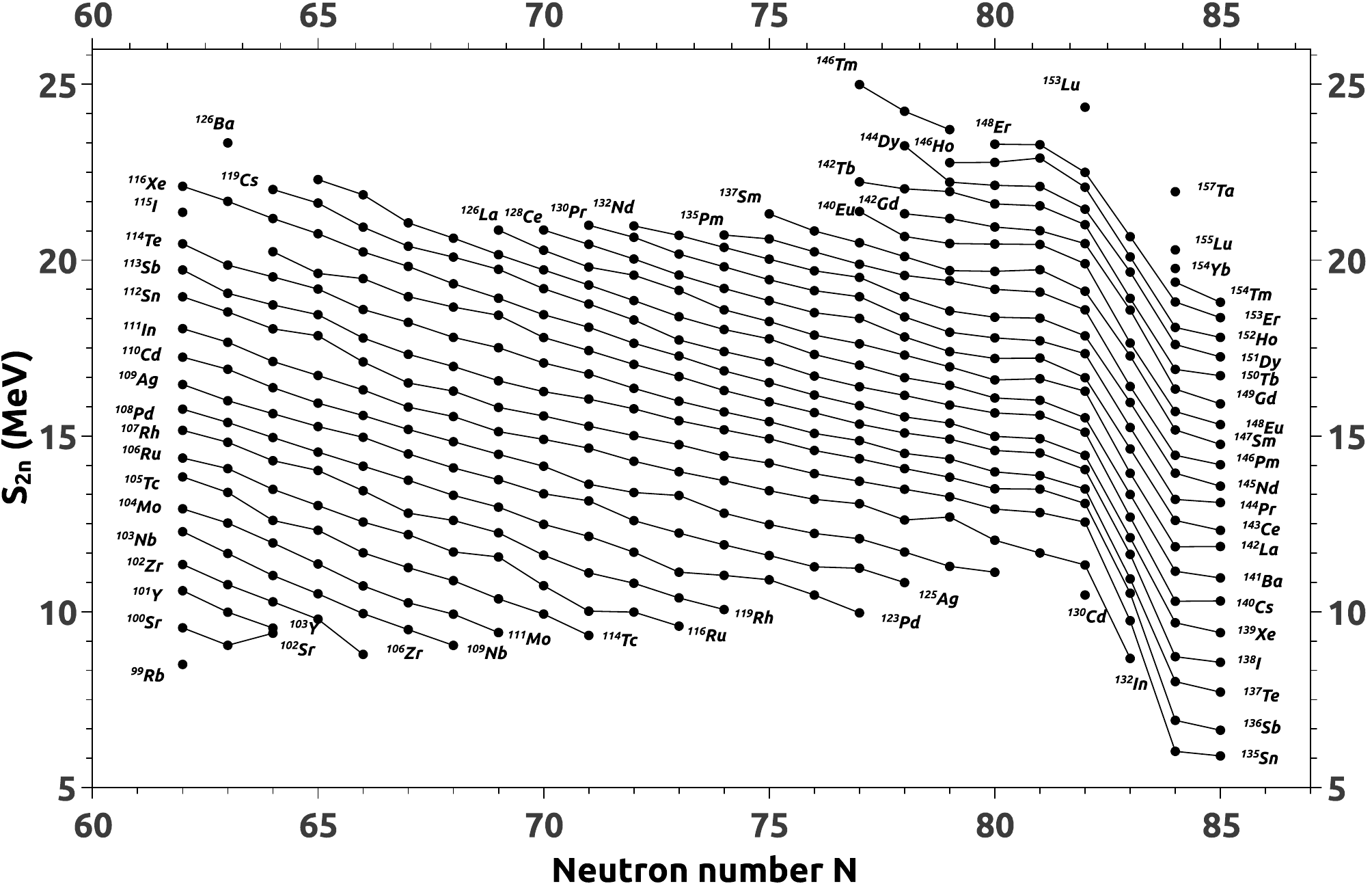}
\caption{A representative S$_{2n}$ mass-sheet before extrapolation.}
\label{Fig:S2nR}
\textcolor{color1}{\rule{\linewidth}{0.72pt}}
\end{figure}

\begin{figure}[htbp!]
\centering
\includegraphics[width=0.9\linewidth]{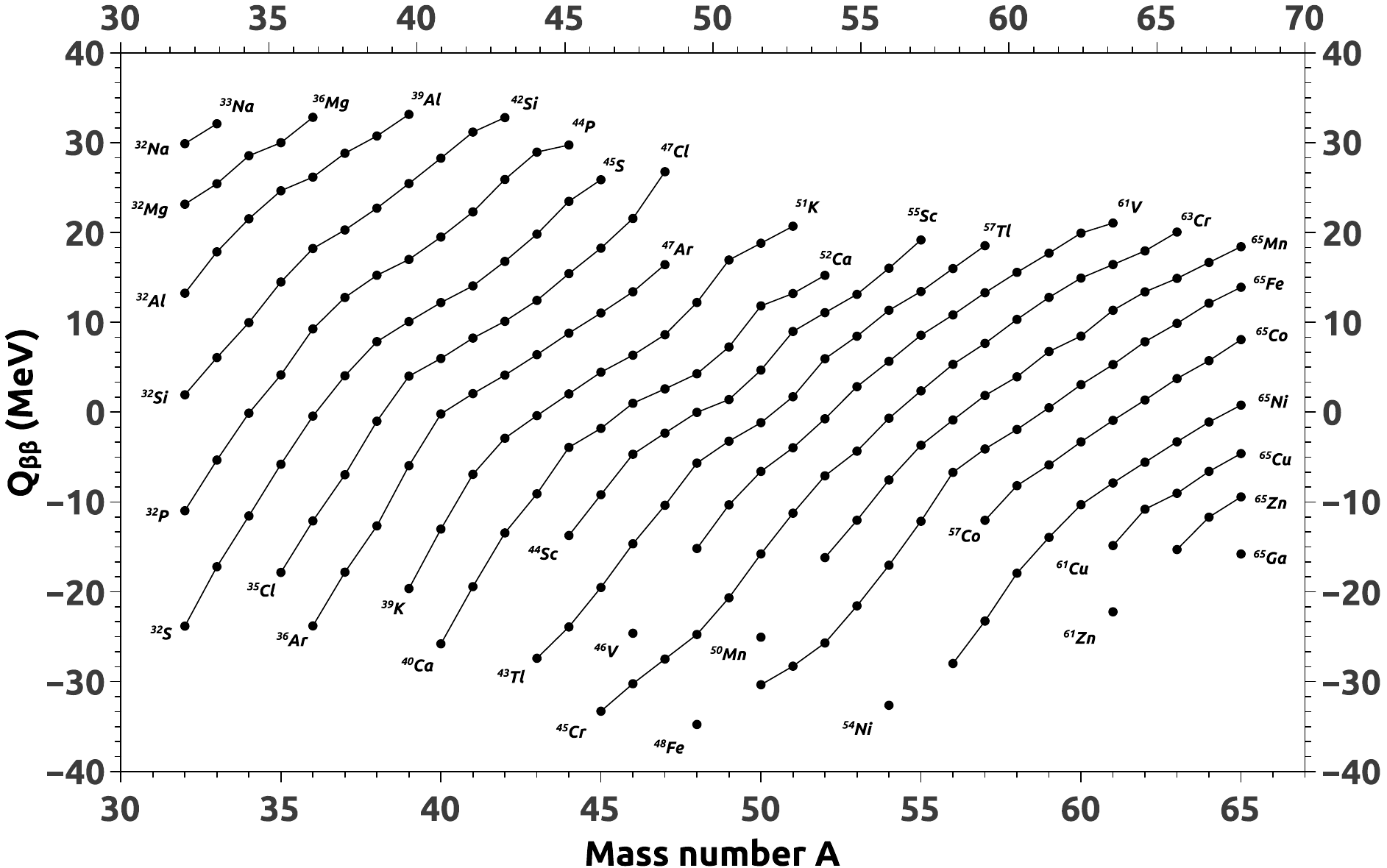}
\caption{A representative Q$_{2\beta^-}$ mass-sheet before extrapolation.}
\label{Fig:Q2bR}
\textcolor{color1}{\rule{\linewidth}{0.72pt}}
\end{figure}

\begin{enumerate}
\item The two-neutron separation energies versus N with lines connecting the isotopes of a given element.
\item The two-proton separation energies versus Z, with lines connecting the isotones.
\item The double-beta decay energies versus A, with lines connecting the isotopes.
\item The four-beta decay energies versus A, with lines connecting the isotopes.
\end{enumerate}
Separation energies in MeV of particles or groups are obtained as the following combination of atomic masses from the AMC12 atomic mass data:
\begin{align} \label{Eq:Eq2}
   S_{2n} &= - M(A, Z ) + M(A - 2, Z ) + 2(n). \\
   S_{2p} &= - M(A, Z ) + M(A - 2, Z - 2) + 2 (^1H).  \\
   Q_{2\beta^-} &= M (A, Z ) - M(A, Z + 2). \\
    Q_{4\beta^-} &= M (A, Z ) - M(A, Z + 4).
\end{align}
Two representative mass surfaces are shown in Figs.~\ref{Fig:S2nR} and~\ref{Fig:Q2bR}, clearly showing the trends in the mass surface. The mass surfaces are calculated using the experimental mass values from the AMC12 mass compilation before extrapolation. Figure~\ref{Fig:S2nR} shows a S$_{2n}$ mass-sheet in which S$_{2n}$ values are plotted as a function of neutron number from 60 to 85. Figure~\ref{Fig:Q2bR} shows a Q$_{2\beta^-}$ mass-sheet in which Q$_{2\beta^-}$ values are plotted as a function of mass number from 32 to 65. The derivative values prior to extrapolation are represented by the filled circles in both graphs. These graphs were used for extrapolation of unknown masses. Precise estimates of unknown masses of nuclides far from the valley of stability have been obtained by extrapolation from well-known mass values on the same sheet.

For example, the two-neutron separation energies, S$_{2n}$(Z, A) have been computed from the ground state nuclear masses M(Z, A) and M(Z, A-2) from the AMC12 and the neutron mass. The evolution of S$_{2n}$ as a function of neutron number shows the well-known regularities (see Fig.~\ref{Fig:S2nR}). For a given proton number, S$_{2n}$ decreases smoothly as the neutron number increases. The curves of various isotopic chains are roughly parallel to each other. Similar calculations for S$_{2p}$, Q$_{2\beta^-}$, and Q$_{4\beta^-}$ were performed using the ground state atomic mass data from AMC12~\cite{Amc14AnndtPfe} for the respective nuclides. The S$_{2p}$ energies versus neutron number, Q$_{2\beta^-}$ energies versus mass number, and Q$_{4\beta^-}$ energies versus mass number have been plotted. The points of isotopic chains have been joined to show the roughly parallel nature of the lines in all cases. All the four sets of plots have been subjected to the extrapolation procedures described here under. In the plots, the experimental data calculated from AMC12 is shown as darkened circles, while the extrapolated data is shown as open circles. The plots in the case of S$_{2n}$ are split into eight neutron-number regions consisting of N= 0-25, 22-45, 42-65, 62-85, 82-105, 102-125, 122-145, and 142-156 and are shown in Figs. 3-10. Similarly, the plots of Q$_{2\beta^-}$ are also divided into nine mass-number (A) regions consisting of A = 0-35, 32-65, 62-95, 92-125, 122-155, 152-185, 182-215, 212-245, and 235-257 and are shown in Figs. 11-19. Figures 3-19 are available through Figshare~\cite{EAmc22FigShr}. It is quite interesting to see the beauty and the consistency of the roughly parallel behavior even with respect to the extrapolated data points, thus showing the quality of the extrapolation method followed in the present work.

In order to estimate the unknown mass values, extrapolations were performed by independently following the trends in the four derivative sheets mentioned above. The value of the extrapolated derivative was then used to obtain the previously unknown mass value. It was observed that point-to-point extrapolation has to be taken into account in order to meticulously account for the local trends in the derivative sheets. In each of the sheets, trends from the three previous (or following) derivatives were used to estimate the value of the unknown derivative. To further improve the accuracy of the extrapolated derivative, it was ensured that the local trends were closely tracked by assigning a higher weightage on the trends from the closest neighbors to the derivative that was to be found. The details of the method used to find the extrapolated derivate and the error in the corresponding extrapolated mass value are provided here. The computational technique developed to obtain the extrapolated derivative is illustrated below with the S$_{2n}$ derivatives as an example.
\subsection*{Extrapolation: The weighted slope method of computing the extrapolated S$_{2n}$ values.} 
S$_{2n_i}$ values in Table~\ref{Tab:Tab1} represent the two-neutron separation energies, where we want to extrapolate to find the unknown two-neutron separation energy labeled S$_{2n_{10}}$. For a given nucleus, the table is listed as a function of the neutron number (N$_i$) and proton number (Z$_j$). To obtain the extrapolated S$_{2n}$ value (S$_{2n_{20}}$ in this case) for a nuclide, we use the trend in the sheet by using the existing derivatives from the already known elements. We have considered point-to-point local extrapolation, in order to follow the trends in the derivative sheets very closely. In our method to obtain the extrapolated S$_{2n}$ value for a nuclide, we used the trend in the sheet by using the already known slopes of the line joining the S$_{2n}$ values for the next three nuclides.
\begin{table}[htbp!]
\centering
\begin{tabular}{| l | l | l | l | l | l | l |}
\hline
\rowcolor{color3} 
N$\setminus$Z &  Z$_{j-1}$ & Z$_{0}$       & Z$_{1}$       & Z$_{2}$       & Z$_{3}$       & Z$_{j+1}$ \\
\hline
N$_{i-1}$     &            &               &               &               &               &           \\
\hline
N$_{1}$       &            & S$_{2n_{10}}$ & S$_{2n_{11}}$ & S$_{2n_{12}}$ & S$_{2n_{13}}$ &           \\
\hline
N$_{2}$       &            & S$_{2n_{20}}$ & S$_{2n_{21}}$ & S$_{2n_{22}}$ & S$_{2n_{23}}$ &           \\
\hline
N$_{i+1}$     &            &               &               &               &               &           \\
\hline
\end{tabular}
\caption{Representative S$_{2n}$ derivative sheet.}
\textcolor{color1}{\rule{\linewidth}{0.72pt}}
\label{Tab:Tab1}
\end{table}

We use the trend in the S$_{2n}$ sheet by finding the slopes of the line joining the S$_{2n}$ values for the neutron numbers N$_{2}$ and N$_{1}$ of (in this case) the next three elements. The slopes for these three known sets of values are calculated as

\begin{align} \label{Eq:Eq3}
m_1 &= \frac{S_{2n_{21}}-S_{2n_{11}}}{N_{2}-N_{1}}, \\
m_2 &= \frac{S_{2n_{22}}-S_{2n_{12}}}{N_{2}-N_{1}},  \\
m_3 &= \frac{S_{2n_{23}}-S_{2n_{13}}}{N_{2}-N_{1}}. 
\end{align}

In order to follow the trend with improved accuracy, we have provided more weightage for the slope calculated from the S$_{2n}$ values immediately next to the one we want to find, than for the slope of the others. As a result, the slope m was found as,
\begin{equation} \label{Eq:Eq4}
m = 0.5m_1 + 0.3m_2 + 0.2m_3.
\end{equation}

Higher weightage is provided for m$_1$ so that trends in slopes closest to the nuclei have more influence than if the three slopes were provided with equal weightage. This makes sure that the local variation in the trend lines is followed closely. The c-intercept for the extrapolated value was obtained by substituting the slope, m to the already known S$_{2n}$ value for the nuclide which is, S$_{2n_{20}}$ and rearranging the following equation:
\begin{equation} \label{Eq:Eq5}
S_{2n_{10}} = mN_2 + c,
\end{equation}
to get the c-intercept:
\begin{equation} \label{Eq:Eq6}
c = S_{2n_{10}} - mN_2.
\end{equation}

Thus, we obtain the extrapolated S$_{2n}$ value as,
\begin{equation} \label{Eq:Eq7}
S_{2n_{20}} = mN_1 + c.
\end{equation}

A similar method has been extended to find the unknown S$_{2n}$ value, depending on where in the sheet of derivatives the method has to be used. In all, this method of local extrapolation by weighted slopes has been used in three different forms, in addition to the way in which it is used above. The second case is similar to the one above, but the two-neutron separation energy that has to be found by extrapolation is S$_{2n_{23}}$, with all the other values in the above table known. For this, we again use the same weightage for slope, where the weightage progressively decreases for the slope that is away from the proton number for which the two-neutron separation energy has to be found. Once the weighted slope is found, the c-intercept is found using the known two-neutron separation energies of the next neutron number (S$_{2n_{13}}$ in this case) and the slope. From the calculated slope and the c-intercept, we then find S$_{2n_{23}}$.

The other two cases are the ones where, for a given neutron number, the two-neutron separation energies of the nuclei with a proton number one higher and one lower are already known. In this case, we perform interpolation to find the two-neutron separation energy using a calculated slope. We first calculate the slope m$_1$ by using the two-neutron separation energies with a proton number one lower and with a neutron number the same as that of the nuclei for which the two-neutron separation energy has to be found, and the neutron number one higher. Similarly, we find the slope m$_2$ from the two nuclei with proton number one higher and the neutron number one lower, and with a neutron number the same as that of the nuclei for which the two-neutron separation energy has to be found. Then the c-intercept is found by using the average of these slopes and the two-neutron separation energy of the nuclei with the same proton number as that for the nuclei for which the two-neutron separation energy has to be found but with a neutron number one lower. Thus, we can interpolate to find the unknown two-neutron separation energy by knowing the slope and the c-intercept as before. The last case is the one that is similar to the one described above, but we also know the two-neutron separation energies of the nuclei with a neutron number lower than the nuclei for which the two-neutron separation energy has to be found and the proton number is lower than, same as, and higher than the nuclide for which the two-neutron separation energy has to be found. That is, we know the two-neutron separation energy of all the nuclei that are around the one for which we need to interpolate. For this case, we not only find the two-neutron separation energy as we did for the previous case but also find the two-neutron separation energy by using the known values with a neutron number that is the same as, and one lower than, the nuclei for which the two-neutron separation energy has to be found. Then we take the average of the two interpolated two-neutron separation energies. 

Similarly, we use these methods to extrapolate (and interpolate) the corresponding values in the S$_{2p}$, Q$_{2\beta^-}$, and Q$_{4\beta^-}$ sheets.
\subsection*{Calculation of errors.} 
The error in mass is a very important quantity to know the validity of the mass value that has been determined by extrapolation. The value of the extrapolated derivative relies on the previously known mass value from which the known derivatives were calculated. The error in the previously known mass value is used to obtain the error in the extrapolated derivatives. The steps involved in determining the error in the extrapolated mass value are described below.

The error in the extrapolated S$_{2n}$, for example, S$_{2n_{10}}$ is calculated from the method used for extrapolation, where we had
\begin{equation} \label{Eq:Eq8}
S_{2n_{10}} = mN_2 + (S_{2n_{20}} - mN_1) = m (N_2 - N_1) + S_{2n_{20}},
\end{equation}
and $\mid$N$_2$ $-$ N$_1\mid$ = 1.

The error in m, the slope, $\Delta$m is calculated using errors in already existing S$_{2n}$ values, used for the calculation of the slope:
\begin{equation} \label{Eq:Eq9}
\Delta m = 0.5(\Delta S_{2n_{11}} + \Delta S_{2n_{21}}) + 0.3(\Delta S_{2n_{12}} + \Delta S_{2n_{22}}) + 0.2(\Delta S_{2n_{12}} + \Delta S_{2n_{22}}),
\end{equation}
where, the error in the known S$_{2n}$ value is $\Delta$S$_{2n}$, is
\begin{equation} \label{Eq:Eq10}
\Delta S_{2n} = \Delta M (A, Z ) + \Delta M (A - 2, Z ) + 2\Delta M(n).
\end{equation}
Where, $\Delta$M's, here represents the errors in the known mass values. Thus, the error in extrapolated S$_{2n}$, $\Delta$S$_{2n_{10}}$ is
\begin{equation} \label{Eq:Eq11}
\Delta S_{2n_{10}} = \Delta m + \Delta S_{2n_{20}}.
\end{equation}
Where, the $\Delta$S$_{2n_{20}}$ is the error in the known derivative value. Then, the error in the extrapolated mass is given by
\begin{equation} \label{Eq:Eq12}
\Delta M_e (A, Z) = \Delta S_{2n_{10}} + \Delta M (A - 2, Z) + 2 \Delta M(n).
\end{equation}
The subscript e in $\Delta$M$_e$(A, Z) denotes that it is the error in the \textit{extrapolated} mass value. Similar calculations have been carried out with different methods of extrapolation, as described above, for each individual nuclide. The procedure has been then, applied for S$_{2p}$, Q$_{2\beta^-}$, and Q$_{4\beta^-}$ extrapolated data.

The extrapolated data of S$_{2n}$, S$_{2p}$, Q$_{2\beta^-}$, and Q$_{4\beta^-}$ were used to calculate the respective extrapolated mass values and thus mass excess values, as well as the uncertainties associated with them. We calculated the weighted average of all the available extrapolated mass excess values and the corresponding error for each nuclide using the standard formula for calculating the weighted averages. The data has been arranged in a tabular form in data set 2~\cite{EAmc22FigShr}, with mass number A, the name of the nuclide, the corresponding atomic number Z, and neutron number N in four consecutive columns. This is followed by the weighted average of extrapolated mass excess value and error designated as AMC12-ME, Error in the next two columns. The next two columns contain the extrapolated mass excess values and error corresponding to that from the AME16 mass evaluation data~\cite{Ame17CpcWan} designated as AME16-ME and Error.

In order to provide a comprehensive comparison, we have collected the theoretical mass model values from four popular mass models: the Duflo-Zuker model (DU-ZU)~\cite{Mic95PhRDuf}, the finite-range droplet model (FRDM12)~\cite{NuG16MolAdndt,Nug95MolAdndt}, the Hartree-Fock-Bogoliubov mean field model (HFB21)~\cite{FurExp10PhRGor}, the infinite nuclear matter model (INM)~\cite{MaP12NayAdndt}, the nuclidic mass formula composed of a gross term, an even-odd term and a shell term is as a revised version of the mass formula constructed by Koura et al. (KUTY)~\cite{Nuc05PtpKou} and published in 2000~\cite{Nuc00NpaHir}, and the most recent WS4 model by Wang et al.~\cite{Sur14PlbWan}. These values corresponding to each nuclide are arranged in the next six columns in Table~\ref{Tab:Tab2}, designated as DU-ZU, HFB21, FRDM12, INM, KUTY, and WS4 respectively.

\section*{Data Records}
All the data generated in the study is available at Figshare~\cite{EAmc22FigShr}. In the present work, more than 1100 experimental data-based mass predictions have been made, in most cases with reduced uncertainties when compared to AME16 data. In the majority of cases, our current data agrees with AME16 data, but with greater precision. More than 370 new masses near the driplines with relatively low uncertainties where there has been no experimental mass data are being presented for the first time. We have mass predictions for more than 100 nuclides in the 1$\leq$Z$\leq$15 region where there were only 50 reported values from AME16 data. The majority of the new data comes from highly neutron deficient nuclei in the lower mass range and highly neutron rich nuclides in the heavy mass range.

In data set 2~\cite{EAmc22FigShr}, we have presented our extrapolation data along with the only available exhaustive extrapolation data from AME16. As there is no way of testing our predictions, we tried to test for self-consistency in data set 3~\cite{EAmc22FigShr}, and in data set 4~\cite{EAmc22FigShr} predictability by using the most recent experimental mass measurements. In data set 5~\cite{EAmc22FigShr}, we have considered a typical region of As, Se, Br, Kr, Rb, Sr, and Y to compare our results with two of the available extrapolations. In the same table, we have also compared our data with one of the most recent mass models, WS4. Four of the five considered models have been widely used for a relatively long time, but the INM model is relatively new and has not yet been tested for its predictive power. Thus, the present work is a real test of the predictive power of these models.

\subsection*{Explanations for the data set 2}
\begin{table}[htbp!]
\centering
\begin{tabulary}{\linewidth}{| l | L |}
\hline
A        & Mass number \\
\hline
Nuclide  & Nucleus \\
\hline
Z        & Atomic number \\
\hline
N	     & Neutron number \\
\hline
AMC12 ME & Weighted average of the available extrapolated mass excess values. \\
\hline
Error    & Error in weighted averaged mass excess. \\
\hline
AME16 ME & Extrapolated mass excess values from AME16. \\
\hline
Error    & Error in extrapolated mass excess values from AME16. \\
\hline
DU-ZU    & Mass excess values from the Duflo-Zuker mass model whereby assuming the existence of a pseudo potential smooth enough to do Hartree-Fock variations and good enough to describe nuclear structure, they construct mass formulas that rely on general scaling arguments and on a schematic reading of shell model calculations. \\
\hline
FRDM12   & Mass excess values from the Finite Range Droplet Model where the calculations are based on the finite-range droplet macroscopic and the folded-Yukawa single-particle microscopic nuclear-structure models, which are completely specified the with considerably improved treatment of deformation and fewer approximations. \\
\hline
HFB21    & Mass excess values from the Hartree-Fock-Bogoliubov mean field model (HFB) mass model, labeled HFB-19, HFB-20, and HFB-21, with unconventional Skyrme forces containing t4 and t5 terms, i.e., density-dependent generalizations of the usual t1 and t2 terms, respectively. \\
\hline
INM      & Mass excess values from the infinite nuclear matter model, in which the classical liquid drop is replaced by an INM sphere defining the interacting many-fermionic liquid. \\
\hline
KUTY     & Mass excess values from the nuclidic mass formula composed of a gross term, an even-odd term and a shell term is as a revised version of the mass formula constructed by Koura et al. \\
\hline
WS4      & Mass excess values from the WS4 model, by Wang et al., an improved version of the macroscopic-microscopic mass formula together with the surface diffuseness correction for unstable nuclei, plus radial basis function (RBF) corrections are combined in the WS4 calculations. \\
\hline
\end{tabulary}
\captionlistentry{}
\label{Tab:Tab2}
\end{table}
\newpage
\includepdf[pagecommand={\thispagestyle{fancy}},landscape=true,pages=-,]{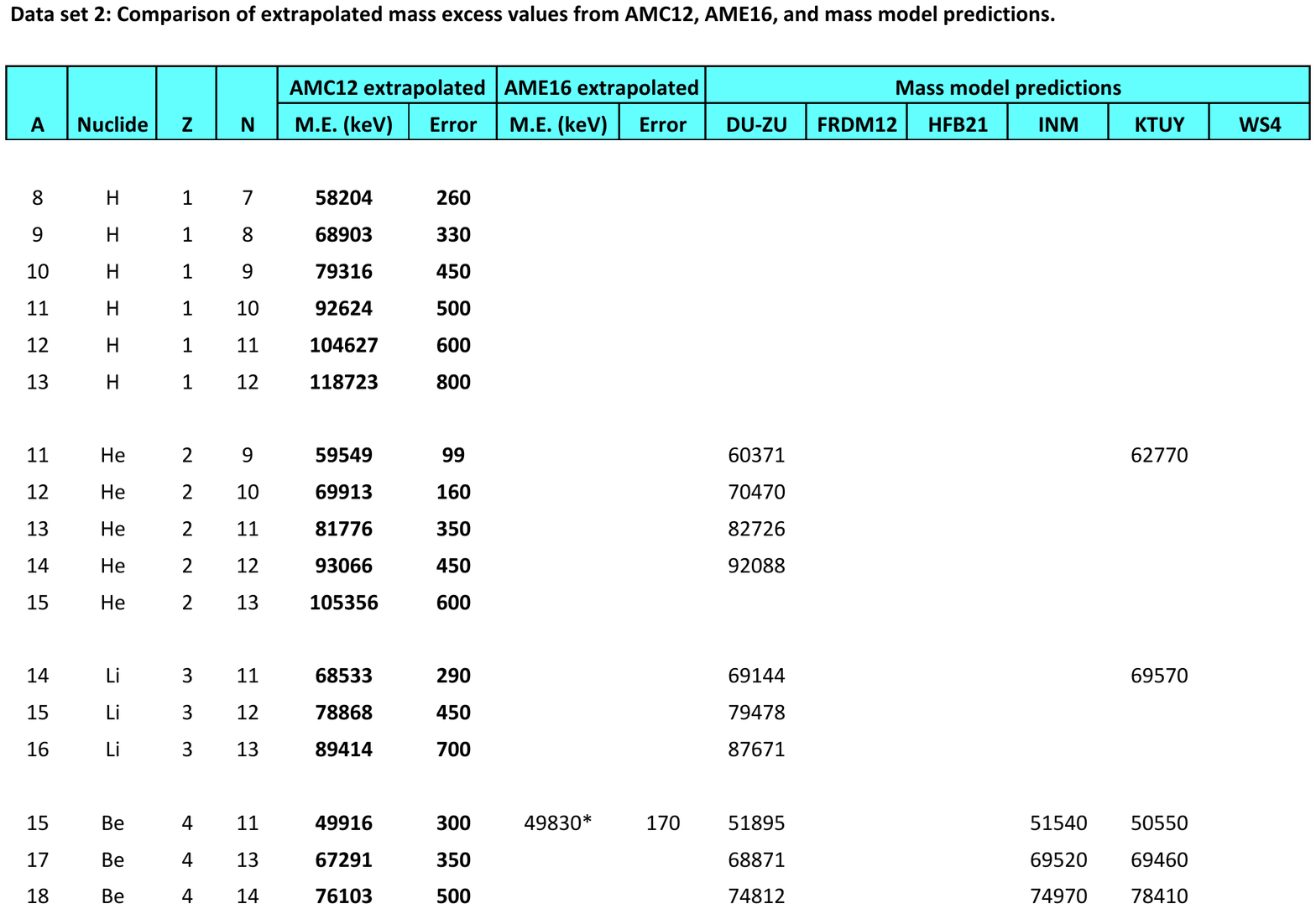}
Data set 2, titled ``comparison of extrapolated mass excess values from AMC12, AME16, and mass model predictions'', is also available through Figshare~\cite{EAmc22FigShr}. The starred (*) M.E. values in the AME16 extrapolated column are the newly measured data reported by AME16 evaluators and are also used in our present work for confirmation of the efficacy of the present extrapolation method in data set 4.

\subsection*{Explanation for data set 3}
\begin{table}[htbp!]
\centering
\begin{tabulary}{\linewidth}{| l | L |}
\hline
A            & Mass number \\
\hline
Z            & Atomic number \\
\hline
N            & Neutron number \\
\hline
Nuclide      & Nucleus \\
\hline
Known        & Calculated S$_{2n}$ value from the AMC12 mass compilation. \\
\hline
Error        & Error in the calculated S$_{2n}$ value. \\
\hline
Extrapolated & The present S$_{2n}$ value extrapolated from the AMC12 mass surface, assuming that it is unknown, using the present weighted slope method. \\
\hline
Error        & The error in the extrapolated S$_{2n}$ value. \\
\hline
Deviation    & Difference between the known S$_{2n}$ value and extrapolated S$_{2n}$ value. \\
\hline
\end{tabulary}
\captionlistentry{}
\label{Tab:Tab3}
\end{table}
\newpage
\includepdf[pagecommand={\thispagestyle{fancy}},landscape=true,pages=-,]{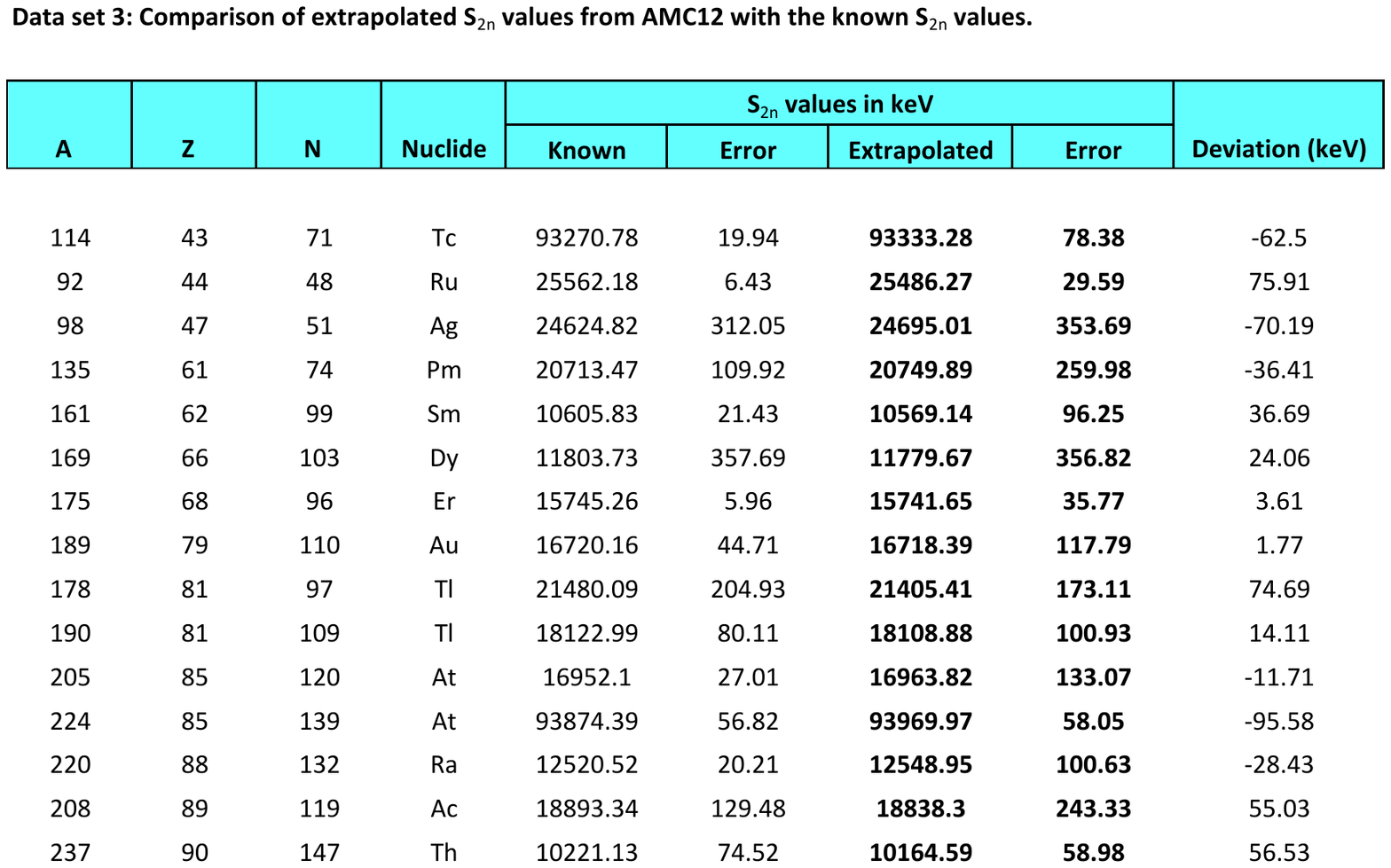}
Data set 3, titled ``comparison of extrapolated S$_{2n}$ values from AMC12 with the known S$_{2n}$ values'', is also available through Figshare~\cite{EAmc22FigShr}.
\subsection*{Explanation for data set 4}
\begin{table}[htbp!]
\centering
\begin{tabulary}{\linewidth}{| l | L |}
\hline
A            & Mass number \\
\hline
Nuclide      & Nucleus \\
\hline
Z            & Atomic number \\
\hline
N            & Neutron number \\
\hline
E.D.-Present & Extrapolated data using the present extrapolation method. \\
\hline
E.D.-AME12   & Extrapolated data from AME12. \\
\hline
New Data     & New experimental data: data not included in AMC12 but reported in AME16 and the most recent (2020) compilation of Kroll et al. \\
\hline
M.E.         & Mass excess value in keV. \\
\hline
Error        & Error in mass excess value. \\
\hline
WS4-M.E.     & Mass excess from WS4. \\
\hline
$\Delta$ME   & The difference between mass excess data. \\
\hline
Present      & The difference between the present extrapolation and the new measurement. \\
\hline
AME12        & The difference between the AME16 extrapolation and the new measurement. \\
\hline
WS4          & The difference between the value from WS4 and the new measurement. \\
\hline
Ref.         & References for the new measurement. \\
\hline
\end{tabulary}
\captionlistentry{}
\label{Tab:Tab4}
\end{table}
\newpage
\includepdf[pagecommand={\thispagestyle{fancy}},landscape=true,pages=-,]{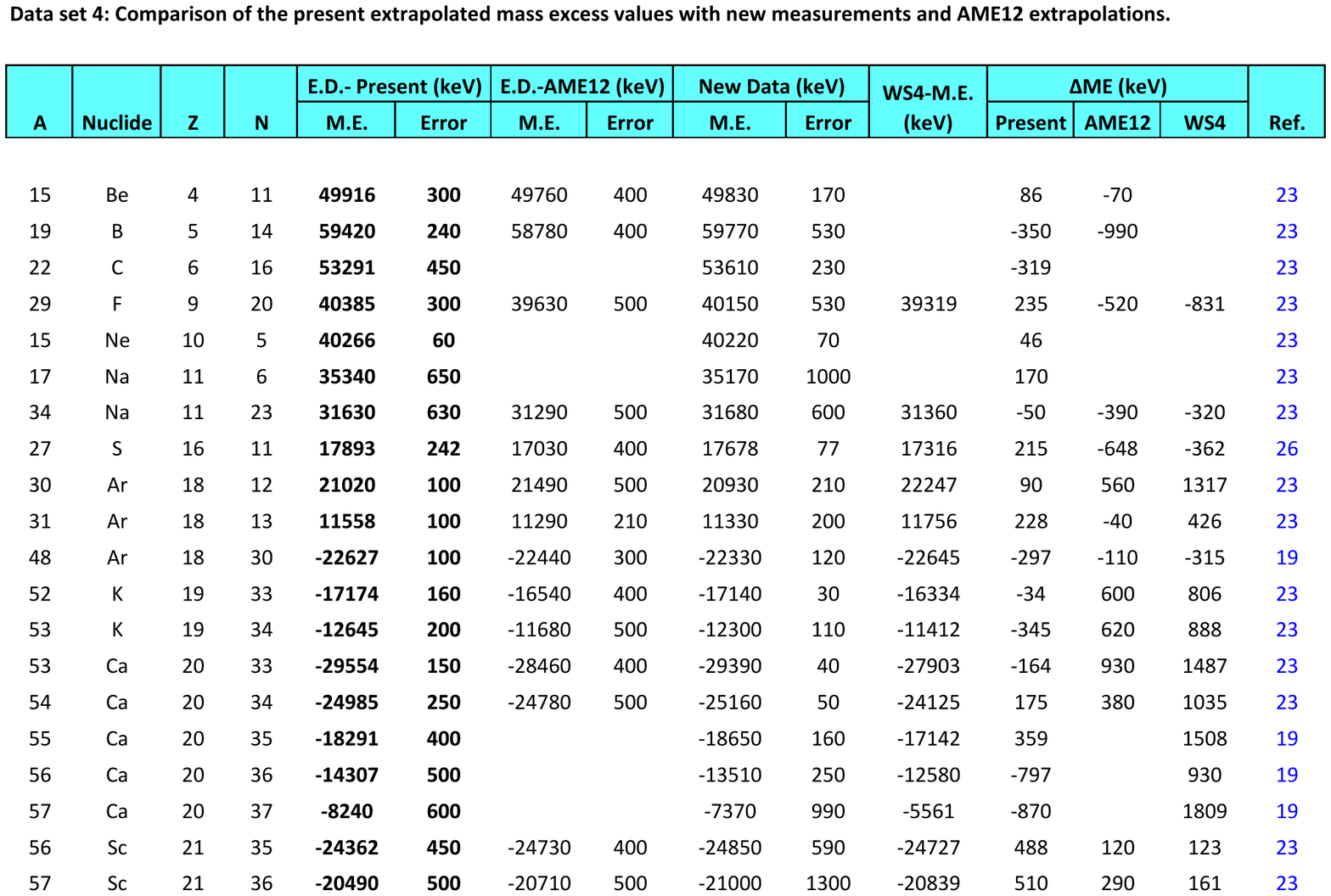}
Data set 4, titled ``comparison of the present extrapolated mass excess values with new measurements and AME12 extrapolations'', is also available through Figshare~\cite{EAmc22FigShr}.
\subsection*{Explanation for data set 5}
\begin{table}[htbp!]
\centering
\begin{tabulary}{\linewidth}{| l | L |}
\hline
A                 & Mass number \\
\hline
Z                 & Atomic number \\
\hline
N                 & Neutron number \\
\hline
Nuclide           & Nucleus \\
\hline
WS4               & The mass excess predictions of the most recent and improved version of the semi-empirical nuclear mass formula, based on the macroscopic method with the Skyrme energy-density functional, together with the surface diffuseness correction for unstable nuclei and the RBF corrections. \\
\hline
P.W.              & Extrapolated mass excess values of the present work using the weighted slope average method. \\
\hline
AME16             & Extrapolated mass excess values from the AME16 mass evaluation. \\
\hline
G.K.               & Extrapolated mass excess values calculated using the Garvey-Kelson relations. \\
\hline
$\Delta_{PW-WS4}$ & Deviation of the present work from the WS4 mass model predictions. \\
\hline
$\Delta_{PW-AME}$ & Deviation of the present work from the AME16 extrapolations. \\
\hline
$\Delta_{PW-G.K}$ & Deviation of the present work from the Garvey-Kelson relations. \\
\hline
\end{tabulary}
\captionlistentry{}
\label{Tab:Tab5}
\end{table}
\newpage
\includepdf[pagecommand={\thispagestyle{fancy}},landscape=true,pages=-,]{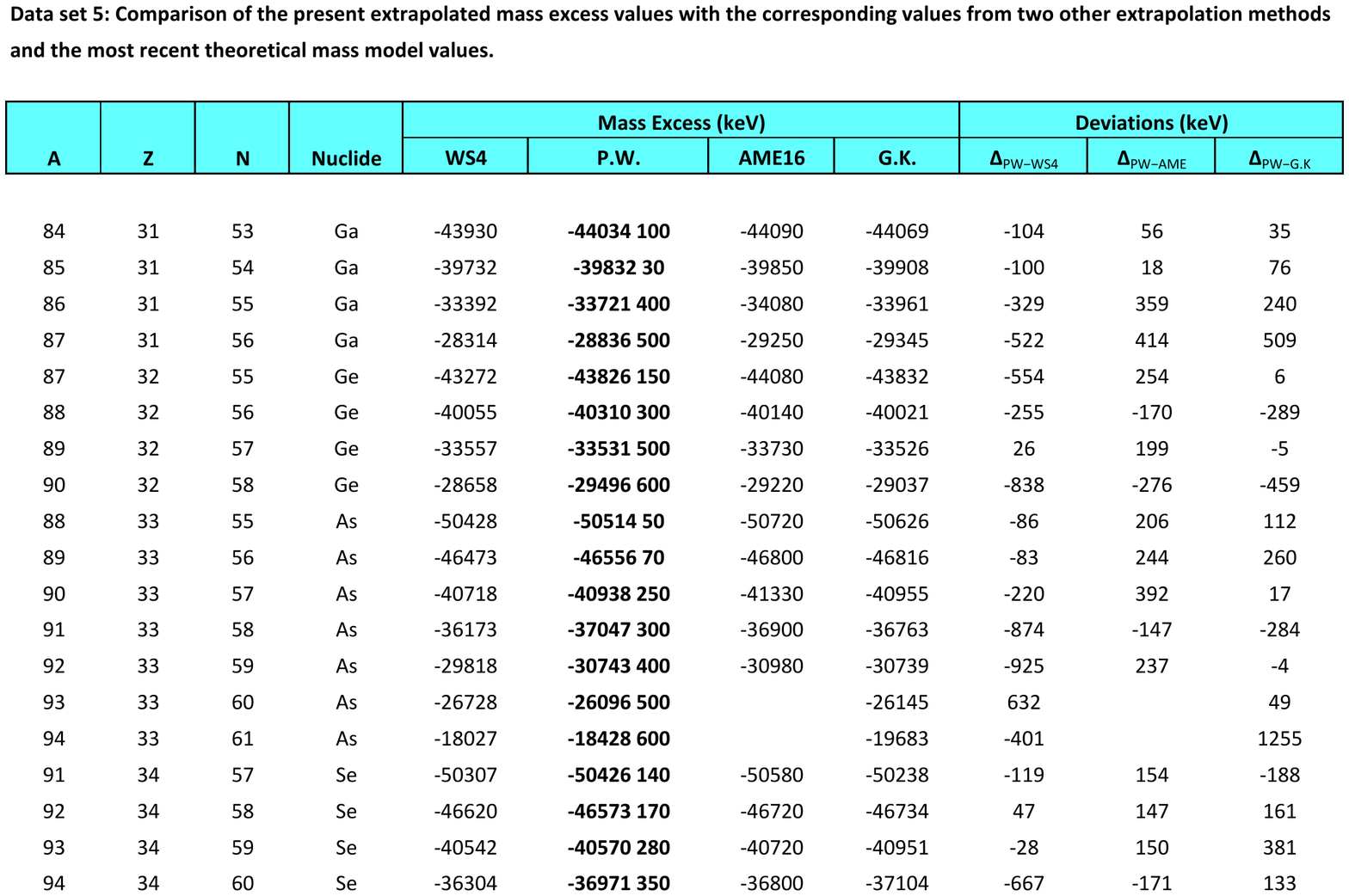}
Data set 5, titled ``comparison of the present extrapolated mass excess values with the corresponding values from two other extrapolation methods and the most recent theoretical mass model values'', is also available through Figshare~\cite{EAmc22FigShr}.
\section*{Technical Validation}
In order to test the efficacy of the present method, we tried to extrapolate for the S$_{2n}$ values of a few known nuclides and compared them with the actual derived values from AMC12 and shown in Table~\ref{Tab:Tab3}. It can be seen that our extrapolation method accurately reproduces the original data. Also, we tried to find how our present extrapolations work in terms of predictability by comparing them with the most recent~\cite{Com20AdndtLia} new mass measurements, which were not included in AMC12~\cite{Amc14AnndtPfe}, here we have also used extrapolated data from AME12~\cite{Ame12CpcWan} for comparisons. Table~\ref{Tab:Tab4} shows how our predicted masses compare well with the newly measured ones~\cite{Bde19PrcSun,Tim16PrcMei,Iso18PrcZha,Mas20PrcRei,New16TepjaKno,Nuc17PrcRou,Mas18PlbXin,Hig19PrcVil,Fir20PrlMan,Mas19PrcXux,Pre18PrlVil,Pre18PrlOrf,Exp20PrcVil,Fir18PrlIto}, which were not included in our evaluations but were extrapolated. These two tables confirm the beauty and efficacy of the new method of extrapolation.

In order to thoroughly evaluate our model, we have run an example test by specifically taking only a subset of known values for As, Se, Br, Kr, Rb, Sr, and Y nuclides for which predictions exist from Garvey and Kelson's global mass formula and the AME16 extrapolations, as well as the most recent WS4 mass model predictions. In in Table~\ref{Tab:Tab5}, we compare our model predictions with these three predictions. Our predictions very well agree with all three extrapolations within our uncertainties.

While the Isobaric Multiplet Mass Equation (IMME) is a powerful interpolation/extrapolation instrument only in the case of lighter nuclei, Garvey and Kelson's global mass formulas will in general be too approximate to make predictions to a sufficient degree of accuracy, and the interactive graphical procedure developed by Audi and Wapstra encompasses a ``subjective'' component in the form of individual judgments. Even though we do not claim superiority of our present local extrapolation method, it is emphasized that the present local extrapolation method does not suffer from the limitations of the earlier methods. From the success of our extrapolation method in predicting the unknown mass values, which can be seen from the comparison with other models, the present method of using weighted slopes to compute S$_{2n}$ values can be said to be an effective method to obtain dependable estimates of unknown, poorly known, or questionable masses, and it also does not depend on any personal judgment.

\section*{Usage Notes}
An attempt has been made to improve the efficacy of local extrapolations of experimental mass data through a simple weighted slope method. The predictive power is remarkably encouraging, as can be seen from the comparison with very recent new mass measurements, where the deviations are well within the experimental uncertainties in most cases. Our extrapolations, in the majority of cases, show very good improvement over the earlier extrapolations.

As can be seen from data set 2~\cite{EAmc22FigShr}, where we have presented our experimental extrapolation data with the only available exhaustive extrapolation data of AME16, our method could predict further away from stability in some cases, and for the majority of the nuclei, the accuracy of our data is either better or comparable. The self-consistency of our method of extrapolation is well tested in data set 3~\cite{EAmc22FigShr} and in data set 4~\cite{EAmc22FigShr}, where we compared our data with the most recent experimental results available, demonstrating the predictability well within the experimental uncertainties. A humble attempt is made in data set 5~\cite{EAmc22FigShr}, where we have considered a typical region of As, Se, Br, Kr, Rb, Sr, and Y and compared our predictions with two available extrapolations and one of the most recent mass models, WS4, we can see the deviations are well within the uncertainties. In our data set 2~\cite{EAmc22FigShr} we have included our extrapolated data, AME16 extrapolated data and the mass model predictions for each nuclide so that the experimentalists can have the full information at hand in one place. This would provide an opportunity for the theoreticians to look for new systematics and trends in different regions of interest and accordingly improve the calculations, while the experimentalists can plan new measurements. Interestingly, with our limited analysis of the comparison with mass model predictions, the most recent WS4 calculations together with the surface diffuseness correction for unstable nuclei and the RBF corrections show very good agreement within the experimental measurements, better than the HFB and Duflo-Zuker formulas. Thus, the extrapolated experimental data of masses can be useful for assessing the impact of current and future experiments in the context of model developments. Additionally, it is anticipated to have an impact on research in fields where experiments are currently not viable, including in simulations of the astrophysical r-process.
\section*{Code availability}
The extrapolation can be performed by a straightforward implementation of the equations provided in the methods section. To aid the implementation of the equations, an example extrapolation is provided in spreadsheet (.xlsx) format in Figshare~\cite{EAmc22FigShr}. The illustration depicts the extrapolation method to obtain the S$_{2n_{10}}$ value in Table~\ref{Tab:Tab1}, and the extrapolated S$_{2n}$ value can then be used to obtain the corresponding unknown mass value. The S$_{2n}$ energies versus neutron number, and the Q$_{2\beta^-}$ energies versus mass number after extrapolation are shown in Figs 3-19 below, and are also available through Figshare~\cite{EAmc22FigShr}. In the plots, the experimental data calculated from AMC12 is shown as darkened circles, while the extrapolated data is shown as open circles.
\newpage
\includepdf[pagecommand={\thispagestyle{fancy}},landscape=true,pages=-,]{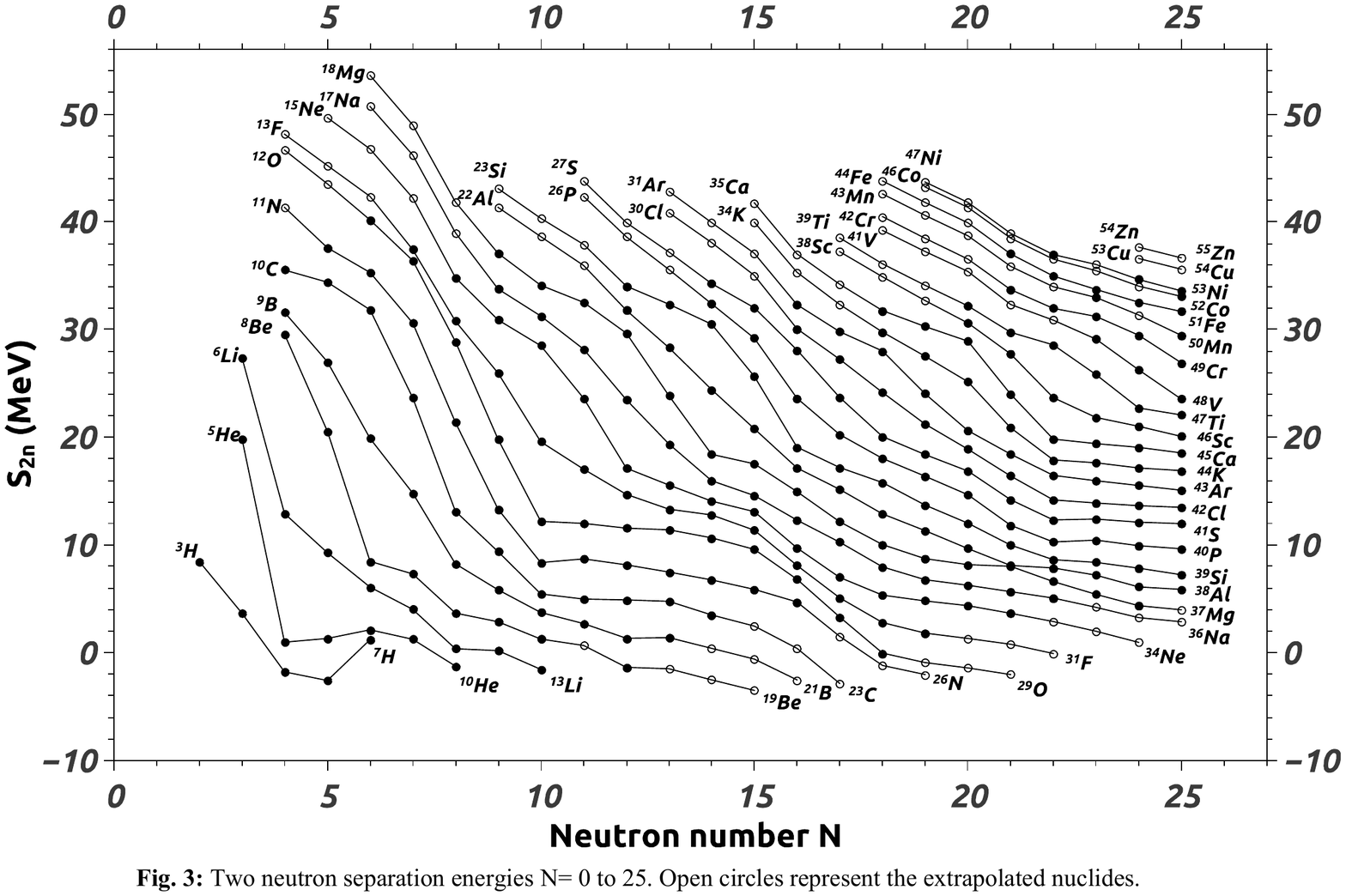}
\bibliography{SciDatBib}

\section*{Author contributions}
K.V. envisioned the extrapolation of new mass values. K.V. and S.C. directed the extrapolation of new mass values. S.R.D.S. developed the extrapolation method, performed the extrapolation, and verified all the data. K.V. performed the weighted averages of mass excess from the four extrapolations and compared the data with other mass models. All authors participated in planning, drafting, and revising the manuscript.

\section*{Competing interests}
The authors declare no competing interests.

\section*{Additional information}
Correspondence and requests for materials should be addressed to S.R.D.S. \\

\noindent
\textcopyright~The Author(s) 2022
\end{document}